\documentclass[pra,aps,groupedaddress,twocolumn,floatfix,nofootinbib]{revtex4}

\usepackage{graphicx}
\newcommand{\beq}{\begin{equation}}
\newcommand{\eeq}{\end{equation}}
\newcommand{\beqa}{\begin{eqnarray}}
\newcommand{\eeqa}{\end{eqnarray}}

\def\half{\frac{1}{2}}
\def\opone{\leavevmode\hbox{\small1\normalsize\kern-.33em1}}

\begin{document}

\title{Time Really Passes, Science Can't Deny That}

\author{Nicolas Gisin}
\affiliation{Group of Applied Physics, University of Geneva, 1211 Geneva 4,    Switzerland}

\date{\small \today}

\begin{abstract}
Today's science provides quite a lean picture of time as a mere geometric evolution parameter. I argue that time is much richer. In particular, I argue that besides the geometric time, there is creative time, when objective chance events happen. The existence of the latter follows straight from the existence of free-will. Following the french philosopher Lequyer, I argue that free-will is a prerequisite for the possibility to have rational argumentations, hence can't be denied. Consequently, science can't deny the existence of creative time and thus that time really passes.
\end{abstract} 

\maketitle

\section{Introduction}
What is free-will for a physicist? This is a very personal question. Most physicists pretend they don't care, that it is not important to them, at least not in their professional life. But if pressed during some evening free discussions, after a few beers, surprising answers come out. Everything from ``obviously I enjoy free-will'' to ``obviously I don't have any free-will'' can be heard. Similarly, questions about time lead to vastly different, though general quite lean discussions: ``Time is a mere evolution parameter'', ``Time is geometrical'' are standard claims that illustrate how poorly today's physics understands time. Consequently, a theory of quantum gravity that will have to incorporate time in a much more subtle and rich way will remain a dream as long as we don't elaborate deeper notions of time. 

I like to argue that some relevant aspect of time is not independent of free-will and that free-will is necessary for rational thinking, hence for science. Consequently, this aspect of time, that I'll name creative time - or Heraclitus-time - is necessary for science. For different arguments in favor of the passage of time, see, e.g., \cite{Norton,Smolin}.

The identification of time with (classical) clocks is likely to be misleading (sorry Einstein). Clocks do not describe our internal feeling of the passage of time, nor the objective chance events that characterize disruptive times - the creative time - when something beyond the mere unfolding of a symmetry happens. Indeed, clocks describe only one aspect of time, the geometric, boring, Parmenides-time. 

But let's start from the beginning. Before thinking of time and even before physics and philosophy, we need the possibility to decide what we'll consider as correct statements that we trust and believe and which statements we don't trust and thus don't buy. Hence:\\ 

Free-Will comes first, in the logical order; and all the rest follows from this premise.\\

Free-will is the possibility to choose between several possible futures, the possibility to choose what to believe and what to do (and thus what not to believe and not to do). This is in tension with scientific determinism\footnote{For physicists, scientific determinism is an extraordinarily strong view: everything is determined by the initial state of the atoms and quanta that make-up the work, nothing beyond that has any independent existence.}, according to which, all of today's facts were necessary given the past and the laws of nature. Notice that the past could be yesterday or the big-bang billions of years ago. Indeed, according to scientific determinism, nothing truly new ever happens, everything was set and determined at the big-bang\footnote{Equally, one may claim that everything is set by tomorrow; a fact that illustrates that time in such a deterministic world is a mere illusion \cite{Barbour}.}. This is the view today's physics offers and I always found it amazing that many people, including clever people, do really believe in this \cite{Barbour}. Time would merely be an enormous illusion, nothing but a parameter labeling an extraordinary unraveling of some pre-existing initial (or final) conditions, i.e. the unfolding of some symmetry. What is the explanatory power of such a view? What is the explanatory power of the claim that everything was set at the beginning - including our present day feelings about free-will - and that there is nothing more to add because there is no possibility to add anything. Clearly, I am not a compatibilist \cite{Kane2005}, i.e. not among those who believe that free-will is merely the fact that we always happen to ``choose'' what was already pre-determined to occur, hence that nothing goes against our apparently free choices\footnote{Compatibilism is quite fashionable among philosophers. They argue that it is our character, reasons and power that determine our actions \cite{Kane2005}. But for a physicist, there is nothing like characters, reasons or power above the physical state of the atoms and quanta that make up our brain, body and all the universe. Hence, if the physical state evolves deterministically, then there is nothing left, everything is determined. In such a case the difference between a human and a laundry machine would only be a matter of complexity, nothing fundamental.}. I strongly believe that we truly make choices among several possible futures.

Before elaborating on all this, let me summarize my argument. The following sections do then develop the successive points of my reasoning.

\section{The logic of the argument}
\begin{enumerate}
\item Free-Will comes first in the logical order. Indeed, without free-will there is no way to make sense of anything, no way to decide which arguments to buy and which to reject. Hence, there would be no rational thinking and no science. In particular, there would be no understanding.
\item Since free-will is the possibility to choose between several possible futures, point 1 implies that the world is not entirely deterministic.
\item Non-determinism implies that time really exists and really passes: today there are facts that were not necessary yesterday\footnote{Admittedly, I use the primitive concepts of {\it today} and {\it yesterday} to get the direction of time, but the existence of creative time is a direct consequence of non-determinism.}, i.e. the future is open.
\item In addition to the geometrical time, there is also creative time. One may like to call the first one Parmenides-time, and the second concept of time Heraclitus-time \cite{ParmenidesHeraclitus}. Both exist.
\item The tension between free-will and creative time on one side and scientific determinism on the other side dissolves once one realizes that the so-called real numbers are not really real: there is no infinite amount of information in any finite space volume, hence initial conditions and parameters defining evolution laws are not ultimately defined, i.e. the real numbers that theories use as inital conditions and parameters are not physically real. Hence, neither Newtonian, nor relativity, nor quantum physics are ultimately deterministic.
\item Consequently, neither philosophy nor science nor any rational argument can ever disprove the existence of free-will, hence of the passage of time.
\end{enumerate}

\section{Free-Will comes first, free-will as a prerequisite for understanding and for science}
As already mentioned in the introduction, free-will comes first. Indeed, free-will is the possibility to choose between several possible futures, like the possibility to choose what to believe and what to do, hence also to choose what not to believe and not to do.

Accordingly, without free-will one could not distinguish truth from false, one could not choose between different views. For example, how could one decide between creationism and Darwinism, if we could not use our free-will to choose among these possibilities? Without free-will all supporters of any opinion would be equally determined (programmed) to believe in their views.

In summary, without free-will there would be no way to make sense of anything, there would be no rational thinking and no science. In particular, there would be no understanding. Furthermore, without free-will one could not decide when and how to test scientific theories. Hence, one could not falsify theories and  science, in the sense of Popper \cite{Popper}, would be impossible.

I was very pleased to learn that my basic intuition, expressed above, was shared and anticipated by a poorly known French philosopher, Jules Lequyer in the 19th century, who wanted to simultaneously validate Science and free-will \cite{Lequyer}. As Lequyer emphasized: ``without free-will the certainty of scientific truths would become illusory''. And (my addition) the consistency of rational arguments would equally become illusory. Lequyer continues: ``Instead of asking whether free-will is certain, let's realize that certainty requires free-will''\footnote{Au lieu de nous demander si la libert\'e est une certitude, prenons conscience que la certitude a pour condition la libert\'e.}.

\begin{figure}
\centering
\includegraphics[width=8cm]{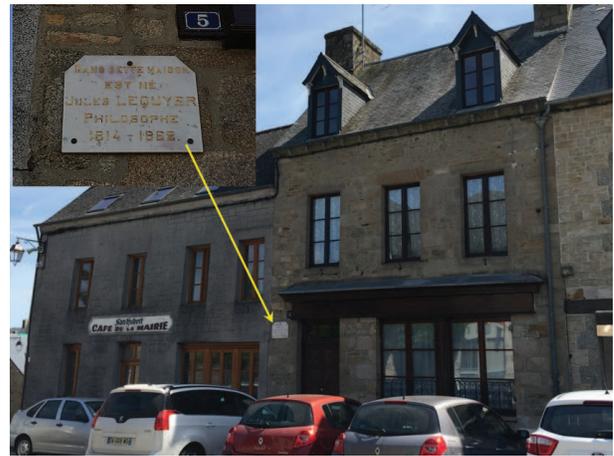}
\caption{Jules Lequyer was born in 1814 in the village Quintin (see inset), in Brittany, France, in this house. He died in 1862, probably committing suicide by swimming away in the sea.}\label{FigLequyer}
\end{figure}

Lequyer also emphasized that free-will doesn't create any new possibilities, it only makes some pre-existing potentialities become actual, a view very reminiscent of Heisenberg's interpretation of quantum theory. However, Lequyer continues, free-will is also the rejection of chance. For Lequyer - and for me - our acts of free-will are beginnings of chains of consequences. Hence, the future is open, determinism is wrong; a point on which I'll elaborate in the next two sections.

Lequyer didn't publish anything. But, fortunately, had an enormous influence on another French philosopher, a close friend, Charles Renouvier who wrote about Lequyer's ideas and published some of Lequyer's notes \cite{Lequyer,Renouvier}. In turn, Renouvier had a great influence on the famous American philosopher and psychologist William James who is considered as one of the most influential American psychologists. William James wrote ``After reading Renouvier, my first act of free-will shall be to believe in free-will''. This may sound bizarre, but, in fact, is perfectly coherent: once one realizes that everthing rests on free-will, then one acts accordingly.

\section{Hence, the World is not deterministic. Reconciling Free-Will with scientific determinism}
The existence of genuine free-will, i.e. the possibility to choose among several possible futures, naturally implies that the world is not entirely deterministic. In other worlds, today there are facts that were not necessary, i.e. facts that were not predetermined from yesterday, and even less from the big-bang. 


Recall that according to scientific determinism everything was set at the beginning, let's say at the big-bang, and since then everything merely unfolds by necessity, without any possible choice.
Philosophers include in the initial state not only the physical state of the universe, but possibly also the character of humans - and living beings. Hence, let's recall that according to physical determinism everything is fully determined by the initial state of all the atoms and quanta at any time (or time-like hypersurface) and the laws of physics. For example, given the state of the universe a nanosecond after the big-bang, everything that ever happened and will ever happen - including the characters, desires and reasons of all humans - was entirely determined by this initial condition. In other words, nothing truly new happens, as everything was already necessary a nanosecond after the big-bang.

But how can one reconcile ideas about free-will such as summarized in the previous sections with scientific determinism? Or even with quantum randomness? This difficulty led many philosophers and scientists to doubt the very existence of free-will. These so-called compatibilist changed the definition of free-will in order to make it compatible with determinism \cite{Kane2005}. Free-will, they argue, is merely the fact that we are determined to never choose anything that doesn't necessary happen. Nevertheless, compatibilists argue, we have the feeling that our ``necessary choices'' are free. This sounds to me like a game of words, some desperate tentative to save our inner feeling of free-will and scientific determinism. But, as Lequyer anticipated, free-will comes first, hence there is no way to rationally argue against its existence, for rational arguing requires that one can freely buy or not buy the argument: genuine compatibilists must freely decide to buy the compatibilists' argument, hence compatibilists must enjoy free-will in Lequyer's sense. Moreover, and this is my main point, scientific determinism is wrong, hence there is no need to squeeze free-will in a deterministic world-view.

Let me emphasize that since free-will comes first, i.e. the possibility to choose between several possible futures comes first, and since this is incompatible with scientific determinism, the latter is necessarily wrong: the future has to be open, as we show in the next section.

Before explaining why physics, including classical Newtonian physics, is not deterministic, let me address first two related questions: When do random (undetermined) events happen? What triggers random events?

Already when I was a high school student, long before thinking seriously about free-will, the concept of randomness and indeterminism puzzled me a lot \cite{GisinPropensity}. When can a random event happen? What triggers its occurrence? If randomness is only a characteristic of long sequences, as my teachers told me, then what characterizes individual random events? What is the probability of a singular event? Aren't long sequences merely the accumulation of individual events\footnote{A long sequence of pseudo-random bits is entirely given at once, because it is entirely determined by the initial condition, i.e. by the seed. In such a case I have no problem with the idea that the pseudo-randomness is a characteristic of the entire sequence. But what about long sequences of truly random bits, produced one after the other, let's say one per second? Each one is a little act of creation and the sequence nothing but an accumulation of individual random bits. Accordingly, randomness of truly random bits must be a characteristic of the individual events, not of the sequence \cite{GisinPropensity}. Notice that in the case of pseudo-randomness only the geometric-boring-time is relevant, but in the case of true randomness that concept of time is insufficient, as the creative-time is at work (but without any free-will).}?

The only interesting answer to the question ``when do random events happen?'' I could find was given by yet another 19th century French philosopher (there is no way to escape from one's cultural environment), Antoine A. Cournot \cite{Cournot}. His idea was that chance happens when causal chains meet. This is a nice idea, illustrated, e.g., by quantum chance which happens when a quantum system encounters a measuring device\footnote{Note that this doesn't solve the quantum measurement problem, i.e. doesn't answer the question ``which configurations of atoms constitute a measurement device?''.}.

This idea can be illustrated by everyday chance events. Imagine that two persons, Alice and Bob meet up by chance
in the street (taken from \cite{Qchance}). This might happen, for example, because Alice was going
to the restaurant further down the same street and Bob to see a friend
who lives in the next street. From the moment they decide to go on foot,
by the shortest possible path, to the restaurant for Alice and to see his
friend for Bob, their meeting was predictable. This is an example of
two causal chains of events, the paths followed by Alice and Bob, which
cross one another and thus produce what looks like a chance encounter
to each of them. But that encounter was predictable for someone with
a sufficiently global view. The apparently chance-like nature of the
meeting was thus only due to ignorance: Bob did not know where Alice
was going, and conversely. But what was the situation before Alice
decided to go to the restaurant? If we agree that she enjoys the benefits
of free-will, then before she made this decision, the meeting was truly
unpredictable. True chance is like this.
True chance does not therefore have a cause in the same sense as events in
a deterministic world. A result subject to true chance is not predetermined in
any way. But we need to qualify this assertion, because a truly chance like
event may have a cause. It is just that this cause does not determine
the result, only the probabilities of a range of different possible results are determined.
In other words, it is only the propensity of a certain event to be realised that is actually predetermined, not which event obtains \cite{GisinPropensity}.

\begin{figure}
\centering
\includegraphics[width=8cm]{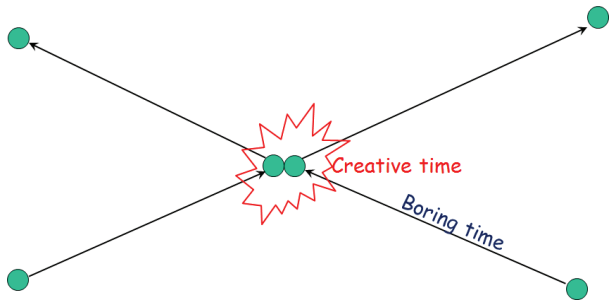}
\caption{Sketch of two colliding classical particles. Initially they merely move along straight lines, nothing happens. Next, they collide, the very detail of this process depends on infinitesimal digits of the initial conditions and of their shapes. Finally, the two particles continue again along boring straight lines.}\label{FigCollidingPart}
\end{figure}

\begin{figure}
\centering
\includegraphics[width=6cm]{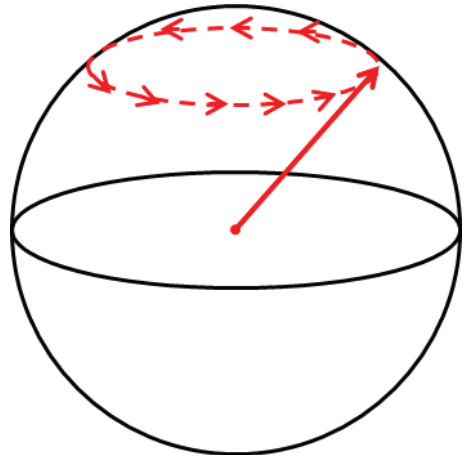}
\caption{Illustration of a unitary evolution in an arbitrary Hilbert space.}\label{FigUnitaryEvol}
\end{figure}

Let's have a more physicist look at that. First, consider two colliding classical particles, see Fig. \ref{FigCollidingPart}. 
Next, consider a unitary quantum evolution in an arbitrary Hilbert space, see Fig. \ref{FigUnitaryEvol}. Look for a while at the latter one; it is especially boring, nothing happens, it is just a symmetry that displays itself. Possibly the symmetry is complex and the Hilbert space very large, but frankly, nothing happens as the equivalence between the Schr\"odinger and the Heisenberg pictures clearly demonstrates. Likewise, for a bunch of classical harmonic oscillators nothing happens. Somehow, there is no time (or only the boring geometric time that merely labels the evolution). Similarly, as long as the classical particles of Fig. \ref{FigCollidingPart} merely move straight at a constant speed, nothing happens: in another reference frame they are at rest. It is only when the classical particles collide, or when the quantum system meets a measuring apparatus, that something happens, as Cournot claimed.

But one may object that in phase space the point that represents the 2 particles doesn't meet anything. In phase space, there is no collision, as collisions require at least two objects and in phase space there is only one object, i.e. one point. Moreover, the collision in real space and the consequence of that collision is already entirely determined by the initial conditions: in phase space it's only a symplectic symmetry that displays itself.

And even if one assumes that each particle is initially “independent”, whatever that could mean, after colliding the 2 particles get correlated. Hence, for Cournot's idea to work, one would need a “correlation sink”. This is a bit similar to the collapse postulate of quantum theory which breaks correlations, i.e. resets independence (separability).

In summary, Cournot's idea is attractive, but not entirely satisfactory; it doesn't seem to fit with scientific determinism. It took me a very long time to realize what is wrong with that claim.

\section{Real numbers are not really real: Mathematical Real Numbers are Physical Random Numbers}
Consider a finite volume of space, e.g. a one millimeter radius ball containing finitely many particles. Can this finite volume of space hold infinitely many bits of information? Classical and quantum theories answer is a clear ``yes''. But why should we buy this assertion? The idea that a finite volume of space can hold but a finite amount of information is quite intuitive. However, theoretical physics uses real numbers (and complex numbers, but let's concentrate on the reals, this suffices for my argument). Hence the question: are so-called real numbers really real? Are they physically real?

For sure, it is not because Descartes (yet another French philosopher, but this time a well-known one) named the so-called real numbers ``real'' that they are really real.

Actually, the idea that real numbers are truly real is absurd: a single real number contains an infinite number of bits and could thus, for example, contain all the answers to all questions one could possibly formulate in any human language \cite{Chaitin}. Indeed, there are only finitely many languages, each with finitely many letters or symbols, hence there are only countably many sequences of letters. Most of them don't make any sense, but one could enumerate all sequences of letters as successive bits of one real number $0.b_1b_2b_3...b_n...$, first the sequences of length 1, next of length 2 and so on. The first bit after each sequence tells whether the sequence corresponds to a binary question and, if so, the following bit provides the answer. Such a single real number would contain an infinite amount of information, in particular, as said, it would contain the answer to all possible questions one can formulate in any human language. No doubt, real numbers are true monsters!

Moreover, almost all so-called real numbers are uncomputable. Indeed, there are only countably many computer programs, hence real numbers are uncomputable with probability one. In other words, almost all real numbers are random in the sense that their sequences of digits (or bits) are random. Let me emphasize that they are as random as the outcome of measurements on half a singlet\footnote{ That is, on a spin $\half$ maximally entangled with another spin $\half$.}. And these random numbers (a better name for “real” numbers) should be at the basis of scientific determinism? Come on, that's just not serious!\\

Imagine that at school you would have learned to name the so-called {\it real numbers} using the more appropriate terminology of {\it random numbers}. Would you believe that these numbers are at the basis of scientific determinism? To name ``random numbers'' ``real numbers'' is the greatest scam and trickery of science; it is also a great source of confusion in the philosophy of science.\\

Note that not all real numbers are random. Some, but only countably many, are computable, like all rational numbers and numbers like $\pi$ and $\sqrt{2}$. Actually, all numbers one may explicitly encounter are computable, i.e. are exceptional.

The use of real numbers in physics, and other sciences, is an extremely efficient and useful idealization, e.g. to allow for differential equations. But one should not make the confusion of believing that this idealization implies that nature is deterministic. A deterministic theoretical model of physics doesn't imply that nature is deterministic.
Again, real numbers are extremely useful to do theoretical physics and calculations, but they are not physically real.

The fact that so-called real numbers have in fact random digits, after the few first ones, has especially important consequences in chaotic dynamical systems. After a pretty short time, the future evolution would depend on the thousandth digit of the initial condition. But that digit doesn't really exist\footnote{It's not that there is a sharp limit on the number of digits, they merely fade off.}. Consequently, the future of classical chaotic systems is open and Newtonian dynamics is not deterministic. Actually most classical systems are chaotic, at least the interesting ones, i.e. all those that are not equivalent to a bunch of harmonic oscillators. Hence, classical mechanics is not deterministic, contrary to standard claims and widely held beliefs.

Note that the non-deterministic nature of physics may leave room for emerging phenomena, like e.g. phenomena that could produce top-down causes, in contrast to the usual down-top causes we are used to in physics \cite{Ellis}. A well-known example of a set of phenomena that emerges from classical mechanics is thermodynamics which can be deduced in the so-called thermodynamical limit. But, rather than going to infinite systems, it suffices to merely understand that classical mechanics is not ultimately deterministic, neither in the initial condition, nor in the set of boundary conditions and potentials required to define the evolution equations.

\begin{figure}
\centering
\includegraphics[width=8cm]{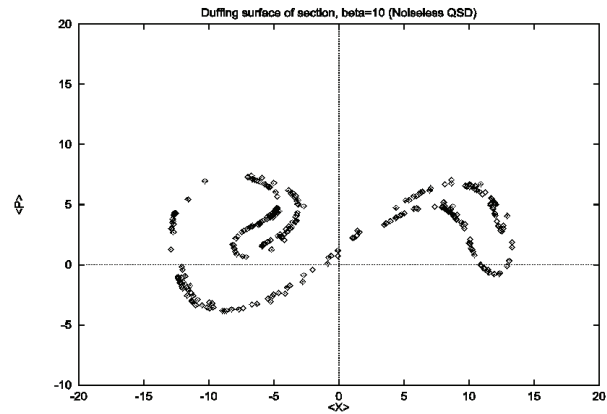}
\caption{Poincar\'e section of the forced and damped quantum Duffing oscillator in the chaotic regime, described by the Quantum State Diffusion model of open quantum systems \cite{QchaosOpenSystem}. Note that the axes represent quantum expectation values of position and momentum. This strange attractor is essentially identical to its classical analog.}\label{FigQchaos}
\end{figure}

What about quantum theory? Well, if one accepts that the measurement problem is a real physics problem - as I do, then this theory is also clearly not deterministic \cite{Qchance}. If, on the contrary, one believes in some form of a many worlds view, then the details of the enormously entangled wave function of the Universe depends again on infinitesimal details, as in classical chaotic systems. Note that although quantum dynamics has no hyper-sensitivity to initial conditions, it shares with classical chaotic systems hyper-sensitivity to the parameters that characterize that dynamics, e.g. the Hamiltonian. Furthermore, open quantum systems recover classical trajectories also in the case of chaotic systems, see Fig. \ref{FigQchaos}. Hence, quantum dynamics is not deterministic. Finally, Bohmian quantum mechanics is again hyper-sensitive to the initial condition of the positions of the Bohmian particles; hence, like chaotic classical systems, Bohmian mechanics is not deterministic.

Admittedly, one may object that now we have an analog of the measurement problem in classical physics, as it is unclear when and how the non-existing digits necessary to define the future of chaotic systems get determined. This is correct and, in my opinion, inevitable. First, because free-will comes first, next because mathematical real numbers are physical random numbers. Finally, because physics - and science in general - is the human activity aiming at describing and understanding how Nature does it. For this purpose one needs to describe also how humans interact with nature, how we question nature \cite{Schrodinger}. Including the observer inside the description results, at best, in a tautology without any possible understanding: there would result no way to freely decide which description provides explanations, which argument to buy or not to buy.

\begin{figure}
\centering
\includegraphics[width=8cm]{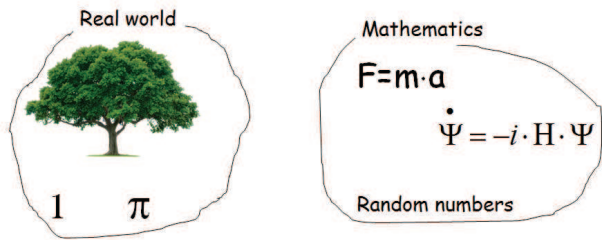}
\caption{The real or physical world versus Pythagoras' mathematical world should not be confused.}\label{FigPhysReal}
\end{figure}

To summarize this section, claiming that classical mechanics is deterministic, or that quantum theory implies a many-world view, is like elevating “real” numbers, the determinism of Newton's equations and the linearity of the Schr\"odinger equation, to some sort of ultimate religious truth. It is confusing mathematics with physics. It is a common but profound epistemological mistake, see Fig. \ref{FigPhysReal}.

\section{Hence, Time really passes. Geometric-boring time versus creative time}
So far we saw that free-will comes first in the logical order, hence all its consequences are necessary. In particular one can't argue rationally against free-will and its natural consequence, namely that time really passes. We also saw that this is not in contradiction with any scientific fact. Actually, quite the opposite, it is in accordance with the natural assumption that no finite region of space can contain more than a finite amount of information. The widely held faith in scientific determinism is nothing but excessive scientism.

This can be summarized with the simple chain of implications:\\

{\it Free-Will} $\Rightarrow$ {\it Non-Determinism} $\Rightarrow$ {\it Time Really Passes}\\

Let us look closer at the implications for time. There is no doubt that time as an evolution parameter exists. To get convinced it suffices to look at a bunch of classical harmonic oscillators (like classical clocks), or the unitary evolution of a closed quantum system, or at the inertial motion of a classical particle as in Fig. \ref{FigCollidingPart}. This time is the boring time, the time when nothing truly new happens, the time when things merely are, time when what matters is {\it being}, i.e. Parmenides-time. One could also name this Einstein's time\footnote{Einstein identified time with classical clocks, i.e. with classical harmonic oscillators. But what about clocks based on Heraclitus' creative time? i.e. clocks based on chaotic or quantum systems?}. But let's look at the collision between the two particles of Fig. \ref{FigCollidingPart}. The detail of the consequences of such a collision depends on non-existing infinitesimal digits, i.e. on mathematically real but physically random numbers. To get convinced just imagine a series of such collisions, this leads to chaos; hence each collision is the place of some indeterminism, that is of some creative time, time when what matters is {\it change}. Hence we call this creative time Heraclitus-time \cite{ParmenidesHeraclitus}. This creative time is extraordinarily poorly understood by today's science, in particular by today's physics. This doesn't mean that it doesn't exist, or that it is not important. On the contrary, it means that there are huge and fascinating open problems in front of us, scientists, physicists and philosophers.

Notice that this is closely related to Cournot's idea that random events happen when independent causal chains meet, e.g. when two classical particles meet. The two particles are independent, at least not fully correlated, because their initial conditions are not fully determined. And their future, after the collision, is not predetermined, but contains a bit of chance.

Similarly, quantum chance happens when a quantum system meets a measurement apparatus, as described by standard textbooks. Admittedly, we don't know what a measurement apparatus is, i.e. we don't know which configurations of atoms constitute a measurement apparatus. This is the so-called quantum measurement problem. According to what we saw, there is a similar problem in classical mechanics: despite the indeterminism in the initial conditions and evolution parameters, things get determined as time passes (as discussed near the end of the previous section).

\section{Conclusion}
Neither philosophy nor science can ever disprove the existence of free-will. Indeed, free-will is a prerequisite for rational thinking and for understanding, as emphasized by Jules Lequyer. Consequently, neither philosophy nor science can ever disprove that time really passes. Indeed, the fact that time really passes is a necessary consequence of the existence of free-will.

The fact that today's science - including classical Newtonian mechanics - is not deterministic may come as a huge surprise to many readers (including the myself of 20 years ago). Indeed, the fact that Descartes named {\it real} numbers that are actually physically {\it random} had enormous consequences. This together with the tendency of many scientists to elevate their findings to some sort of quasi-religious ultimate truth - i.e. scientism - lead to great confusion, as illustrated by Laplace famous claim about determinism and by believers in some form of the many-world interpretation of quantum mechanics, based respectively on the determinism of Newton's equation and on the linearity of Sch\"odinger's equation.


Once one realizes that science is not entirely deterministic, though it clearly contains deterministic causal chains, one faces formidable opportunities. This might seem frightening, though I always prefer challenges and open problems to the claim that everything is solved.

Non-determinism implies that time really passes, most likely at the junction of causal chains, i.e. when creative time is at work. This leaves room for emerging phenomena, like thermodynamics of finite systems. It may also leave room for top-down causality: the initial indeterminism must become determined before indeterminism hits large scale, much in the spirit of quantum measurements.

As a side conclusion, note that robots based on digital electronics will never leave room for free-will, hence the central thesis of hard artificial intelligence (the claim that enough sophisticated robots will automatically become conscious and human-like) is fundamentally wrong.

So, am I a dualist? Possibly, though it depends what is meant by that. For sure I am not a materialist. Note that today's physics already includes systems that are not material in the sense that they have no mass, like electro-magnetic radiation, light and photons. What about physicalism? If this means that everything can be described and understand by today's physics, then physicalism is trivially wrong, as today's theories describe at best 5\% of the content of the universe. More interestingly, if physicalism means that everything can be understood using the tools of physics, then I adhere to this view, though the fact that free-will comes first implies that physics can make endless progress, but without ever reaching a final point. We will understand much more about time and about free-will, though we'll never get a full rational description and understanding of free-will. Just imagine this debate a century ago. How naive anyone claiming at that time that physics provides a fairly complete description of nature would appear today. Similarly, for anyone making today similar claims.

Let me make a last comment, a bit off-track. Free-will is often analyzed in a context involving human responsibility, ``How could we be responsible for our actions if we don't enjoy free-will?''. There is another side to this aspect of the free-will question: ``How could we prevent humans from destroying humanity if we claim we are nothing more than sophisticated robots?'', and ``How could one argue that human life has some superior value if we pretend we are nothing but sophisticated robots?''.\\

\small

\section*{Acknowledgment} This work profited from numerous discussions, mostly with myself over many decades during long and pleasant walks. I should also thank my old friend Jean-Claude Zambrini for introducing me to Cournot's idea, when we were both students in Geneva. Thanks are due to Chris Fuchs who introduced me to Jules Lequyer and to many participants to the workshop on {\it Time in Physics} organized at the ETH-Zurich by Sandra Rankovic, Daniela Frauchiger and Renato Renner.


\begin{thebibliography}{99}

\bibitem{Norton} J. Norton, www.pitt.edu/~jdnorton/Goodies/passage
\bibitem{Smolin} L. Smolin, {\it Time Reborn}, Houghton Mifflin Harcourt, 2013.
\bibitem{Barbour} J. Barbour, {\it The End of Time}, Oxford Univ. Press, 1999.
\bibitem{Kane2005} R. Kane, {\it A contemporary introduction to Free Will}, Oxford Univ. Press, 2005.
\bibitem{ParmenidesHeraclitus} Ch.H.S Temann and Ch.H. Sötemann, {\it Heraclitus and Parmenides - an ontic perspective}, GRIN Verlag GmbH, 2013; 
 K. Jaspers, {\it Anaximander, Heraclitus, Parmenides, Plotinus, Laotzu, Nagarjuna}, Mariner Books, 1974.
\bibitem{Popper}  K.R. Popper, {\it Logik der Forschung}, 1934; {\it The Logic of Scientific
Discovery}, Hutchinson, London, 1959.
\bibitem{Lequyer} J. Lequier, {\it Comment trouver, comment chercher une v\'erit\'e premi\`ere}, Edition de L'Eclat, 1985; see also J. Grenier, {\it La philosophie de Jules Lequier}, Calligrammes, 1983 (ISBN 2 903258 30 9); J. Wahl, {\it Jules Lequier}, Editions des Trois Collines, Gen\`eve-Paris, 1948.
\bibitem{Renouvier} W. Logue {\it Charles Renouvier, Philosopher of Liberty}, Louisiana State Univ. Press, 1993.
\bibitem{GisinPropensity} N. Gisin, {\it Propensities in a non-deterministic physics}, Synthese {\bf 98}, 287 (1991).
\bibitem{Cournot} A. Cournot, {\it Exposition de la th\'eorie des chances et des
        probabilit\'es}, Librairie Hachette,  1843. Reprinted in part in {\it
        Etudes pour le centenaire de la mort de Cournot}, ed. A. Robinet,
        Edition Economica, 1978.
\bibitem{Qchance} N. Gisin, {\it Quantum Chance, nonlocality, teleportation and other quantum marvels}, Springer, 2014.
\bibitem{Chaitin} G. Chaitin, The Labyrinth of the Continuum, in Meta Math!, Vintage, 2008.
\bibitem{Ellis} G.F.R. Ellis, in {\it Downward Causation and the Neurobiology of Free Will}, eds Murphy, Ellis and O'Connor, Springer, 2009.
\bibitem{QchaosOpenSystem} T.A. Brun, N. Gisin et al., Phys. Lett. A 229, 267 (1997).
\bibitem{Schrodinger} E. Schr\"odinger, Mind and Matter, Cambridge Univ. Press, 1958.


\end{thebibliography}
\end{document}